# A Fractional-Order Nonlinear Backstepping Controller Design for Current-Controlled Maglev System

**Dorukhan ASTEKİN[1]*, Fatih ADIGÜZEL[1]**

*1 Yıldız Technical University, Electric-Electronic Faculty, Department of Control and Automation Engineering, İstanbul, Turkey*

** dorukhan.astekin@std.yildiz.edu.tr*

**Abstract**

The magnetic levitation system (Maglev) is a nonlinear system by which an object is suspended with no support other than magnetic fields. The main control perspective of the Maglev system is to levitate a steel ball in air by the electromagnetic force. However, the Maglev system has highly nonlinear dynamics which is inconvenient in the sense of sensitive control/regulation of its nonlinear dynamics. In this paper, the nonlinear backstepping controller based on the fractional-order derivative is proposed for the control of the nonlinear current-controlled Maglev system. After, the system dynamics and fractional-order backstepping controller design are given, the asymptotic stability of the closed-loop system is proved by employing the Lyapunov theory. Some computer-based numerical experiments are carried out to show the effectiveness of the proposed controller for the control of Maglev system.

## 1. Introduction

Fractional (non-integer) calculus is one of the oldest branches of mathematics, and its application to many dynamic systems around us has become a phenomenon. Such systems are expressed by differential equations and applications in control theory, where the fractional calculus concept is widely used, include observer design, system identification and discrete-time systems [1-4]. Studies in nonlinear control systems, which is one of the areas where this concept is extensively used, are increasing rapidly day by day. In nonlinear control systems, as the order of the system increases, the structure and solution of the system becomes more complex. If these systems are combined with fractional-order or complex-order systems, they can become real life applications. By doing so, the error difference between the theoretical and practical application is further eliminated. The application of fractional-order derivative is often based on rationalizing the selected optimal method and the aim is to use fractional-order control (FOC) to improve the performance and efficiency of the system.

Magnetic levitation systems, the most common type of levitation, are the process of counteracting gravity with a magnetic field. In these systems, especially in control theory, many controllers have been designed to make the system asymptotic stable. Some of the studies related to this subject can be listed as follows. In [5], a backstepping controller is designed based on a nonlinear system model in the presence of parameter uncertainties. In [6], a fractional-order controller is designed for an integer-order dynamic system with an adaptive backstepping control approach by considering model uncertainties and external disturbances. Related to this issue at [7], an adaptive fractional backstepping controller is designed for integer-order nonlinear uncertain systems of arbitrary order with prescribed performance subject to unknown time-varying disturbances. In [8], an adaptive non-singular terminal sliding mode controller based on disturbance compensation technique is proposed to control the position of the magnetic levitation system. In [9], a new PI-PD controller design procedure for magnetic levitation system using weighted geometrical center method is presented. In [10], a combination of a robust backstepping controller and an integral action for a magnetic levitation system is presented. In [11], the design and application procedure of an adaptive backstepping controller for magnetic levitation systems is presented. In [12], an implementation of a fractional-order PI controller in a magnetic levitation system is investigated. In [13], a digital fractional-order PID controller is designed to control the position of the levitated object in a magnetic levitation system and compared with an integer-order PID controller. In [14], a design method for tuning fractional-order controllers for second-order unstable processes using stability analysis for the magnetic levitation system is presented. In [15], a solution consisting of a special type of fractional-order controller for the stabilization of second-order unstable plants in a magnetic levitation system is given.

In this paper, a nonlinear fractional backstepping controller is addressed for the current-controlled magnetic levitation system. As different from the classical backstepping controller, the controller utilized in this paper is constructed with the help of the fractional-order derivative instead of the classic derivative operator. In this way, a fractional nonlinear controller is designed, and the asymptotic stability analysis of the closed-loop system is proved in the sense of Lyapunov. Some numerical simulation tests are performed to show the viability of the designed controller for the control of the Maglev system.

The rest of this article is organized as follows. In the next section, the nonlinear model of magnetic levitation is given. After this section, the proposed design controller is presented with the stability analysis in the sense of Lyapunov theory. In Section 4, the effectiveness of the designed controller for the current-controlled Maglev is tested via numerical simulations. The paper ends with some conclusions in Section 5.

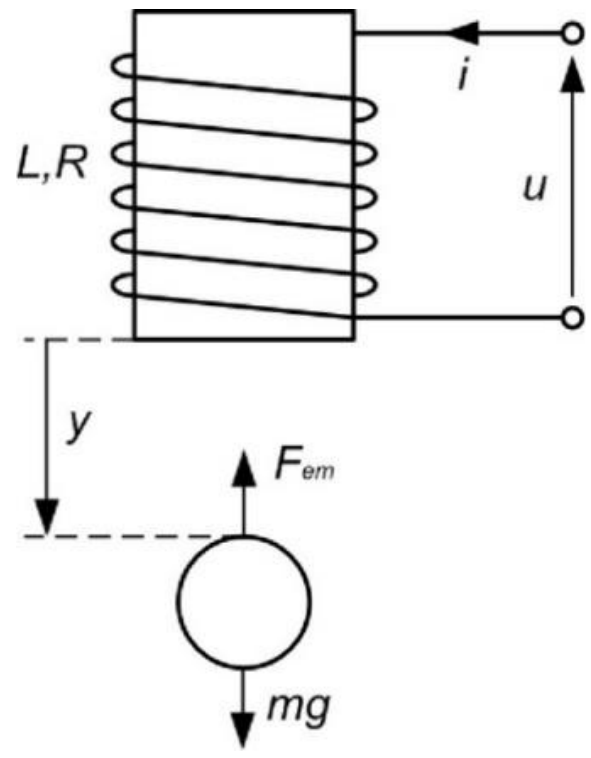

**Fig. 1** The diagram of Maglev system

## 2. Model of Maglev System

In magnetic levitation system, a changeable electromagnetic field occurs through a coil, and a metal ball can be suspended with the help of the produced electromagnetic field. In order to control of the position of the metal ball in advanced, a well-structured controller is need. The diagram of Maglev system is presented in Fig. 1, where $y$, $v$ and $i$ denote the distance between the ball and the coil, the velocity of the ball on the vertical axis, and the coil current, respectively. The parameter $R$ stands for the coil resistance. The parameter $L$ is the coil inductance, $m$ and $g$ present the mass of the ball and the gravitational acceleration, respectively. $F_{em}$ is the electromagnetic force generated by the coil. In the Maglev system, the coil current is adjusted by a current feedback power amplifier as in the many literature studies [5], [9], [16], and in this way the dynamics of the Maglev system is formulated as follows:

$$\frac{dy(t)}{dt} = v(t) \tag{1}$$

$$\frac{dv(t)}{dt} = g - \frac{F_{em}(t)}{m} u(t) \tag{2}$$

where: the following dynamic equation expresses the electromagnetic force $F_{em}$ generated by the coil.

$$F_{em}(t) = \frac{Q}{2m\left(Y_{\infty} + y(t)\right)^2} \tag{3}$$

where: $Q$ and $Y_{\infty}$ are positive system constants and their values are changed by the process of manufacture of the coil, magnetic core, and the metal ball. On the other hand, there are definitions such as Riemann-Liouville, Grünwald-Letnikov, Hölder, Caputo for the definition of fractional-order derivative in the literature. In this study, an application is performed based on the Caputo definition for the fractional-order derivative operator. For a differentiable function $f$, the definition of Caputo fractional-order derivative is defined as in (4) [17].

$$_0D_t^{-\alpha} f(t) = \frac{1}{\Gamma(\alpha - n)} \int_0^t \frac{f^{(n)}(\tau)}{(t-\tau)^{\alpha+1-n}} d\tau, \quad \alpha \in (n-1, n) \tag{4}$$

where: $\Gamma(n)$ is the gamma function, and is described by the following integral:

$$\Gamma(n) = \int_0^{\infty} t^{n-1} e^{-t} dt \ . \tag{5}$$

Furthermore, the Gamma function satisfies the following functional equation:

$$\Gamma(n+1) = n!. \tag{6}$$

## 3. Controller Structure

This section presents the design procedure of a fractional-order backstepping controller. In the model described above, let us define $x_1 = y$, $x_2 = v$, $u(t) = i^2(t)$ and $\phi(x_1) = F_{em} / m$. Thus, the Maglev system can be transformed as equation (7) and

(8).

$$\dot{x}_1 = x_2 \tag{7}$$

$$\dot{x}_2 = g - \phi(x_1)u. \tag{8}$$

Hence, considering the system dynamics (1)-(2), the error dynamics can be given as follows:

$$e_1 = x_1 - x_{1d} \Rightarrow \dot{e}_1 = \dot{x}_1 - \dot{x}_{1d} = x_2 - x_{2d} \tag{9}$$

$$e_2 = x_2 - x_{2d} \Rightarrow \dot{e}_2 = \dot{x}_2 - \dot{x}_{2d} = g - \phi(x_1)u - \dot{x}_{2d} \tag{10}$$

where: $x_{1d}$ denotes the reference trajectory, and $x_{2d} = \dot{x}_{1d}$. Utilizing the equations (9) and (10), the following relationship can be obtained:

$$\dot{e}_1 = e_2 \tag{11}$$

To provide the stability of the dynamic given in (11), the candidate Lyapunov function can be defined as:

$$V_1 = |e_1| \Rightarrow \dot{V}_1 = |\dot{e}_1| = \dot{e}_1 \text{sgn}(e_1) = e_2 \text{sgn}(e_1). \tag{12}$$

For the backstepping procedure, the following the dynamics of $z_2$ can be defined:

$$z_2 = e_2 - \phi_2. \tag{13}$$

Afterwards,

$$\dot{V}_1 = (z_2 + \phi_2)\text{sgn}(e_1) \tag{14}$$

is obtained, and the virtual controller signal can be designed as

$$\phi_2 = -k_1 e_1. \tag{15}$$

Moreover, we consider ${}_0D_t^{-\alpha}|e_1(t)|$ as a fractional-order operator as different from the classical backstepping control method. Then, the dynamics of $z_2$ is analyzed as follows:

$$z_2 = -k_2 \text{sgn}(e_1) - k_3 \text{sgn}(e_1)\, {}_0D_t^{-\alpha}|e_1(t)|. \tag{16}$$

Utilizing (16), (14) becomes

$$\dot{V}_1 = \left(-k_1 e_1 - k_2 \text{sgn}(e_1) - k_3 \text{sgn}(e_1)\, {}_0D_t^{-\alpha}|e_1(t)|\right)\text{sgn}(e_1) \tag{17}$$

$$\begin{aligned} \dot{V}_1 &= -k_1|e_1| - k_2 - k_3\, {}_0D_t^{-\alpha}|e_1(t)| \\ &\le -k_1|e_1| - k_2 - k_3 L_1 = -k_1|e_1(t)| - \gamma_1 \le -\kappa_1\|Y\| \end{aligned} \tag{18}$$

where: ${}_0D_t^{-\alpha}|e_1(t)| \ge L_1$, $\gamma_1 = k_2 + k_3 L_1$ and $\kappa_1 = \min(k_1, \gamma_1)$. Another candidate Lyapunov function to guarantee the stability of the system can be defined as:

$$V_2 = |e_1| + |z_2|. \tag{19}$$

According to the above candidate Lyapunov function, the dynamics of $z_2$ is formed as follows:

$$\dot{z}_2 = \dot{e}_2 - \dot{\phi}_2 = g - \phi(x_1)u - \dot{x}_{2d} - \dot{\phi}_2. \tag{20}$$

Taking the derivative of the Lyapunov function,

$$\begin{aligned}\dot{V}_2 &= |\dot{e}_1| + |\dot{z}_2| = \dot{e}_1 \operatorname{sgn}(e_1) + \dot{z}_2 \operatorname{sgn}(z_2) \\ &= e_2 \operatorname{sgn}(e_1) + \left(g - \phi(x_1)u - \dot{x}_{2d} - \dot{\phi}_2\right)\operatorname{sgn}(z_2) \\ &= (z_2 + \phi_2)\operatorname{sgn}(e_1) + \left(g - \phi(x_1)u - \dot{x}_{2d} - \dot{\phi}_2\right)\operatorname{sgn}(z_2)\end{aligned} \tag{21}$$

$$\dot{V}_2 = -k_1 e_1 \operatorname{sgn}(e_1) + z_2 \operatorname{sgn}(e_1) + \left(g - \phi(x_1)u - \dot{x}_{2d}\right)\operatorname{sgn}(z_2) - \dot{\phi}_2 \operatorname{sgn}(z_2) \tag{22}$$

is obtained. Eventually, the control signal is defined for the asymptotic stability as follows:

$$u = \frac{1}{\phi(x_1)}\left(g - \dot{x}_{2d} + |z_2|\operatorname{sgn}(e_1) + k_2 z_2 + k_3 \operatorname{sgn}(z_2) + k_4 \operatorname{sgn}(z_2)\, {}_0D_t^{-\alpha}|e_1(t)| - \dot{\phi}_2\right). \tag{23}$$

Plugging the designed controller given in (23) into (22) gives

$$\begin{aligned}\dot{V}_2 &= -k_1|e_1| - k_2|z_2| - k_3 - k_4\, {}_0D_t^{-\alpha}|e_1(t)| \\ &\leq -k_1|e_1| - k_2|z_2| - k_3 - k_4 L_1 \leq -k_1|e_1| - k_2|z_2| - \gamma_2 \leq -\kappa_2\|Y\|\end{aligned} \tag{24}$$

where: ${}_0D_t^{-\alpha}|e_1(t)| \geq L_1$, $\gamma_2 = k_3 + k_4 L_1$ and $\kappa_2 = \min(k_1, k_2, \gamma_2)$.

## 4. Simulation Results

To show the effectiveness and viability of the fractional-order backstepping controller design for the control of current controlled magnetic levitation system, several numerical simulations are carried out using MATLAB. In these simulations, the solver step time for the Maglev system was set to 0.1 ms, while the controller sampling period was chosen as 1 ms. The nominal parameters of Maglev system are $m = 0.005$kg, $R = 22\Omega$, $L = 0.5H$, $g = 9.81 m/s^2$, $Q = 0.003$ and $Y_\infty = 0.041$.

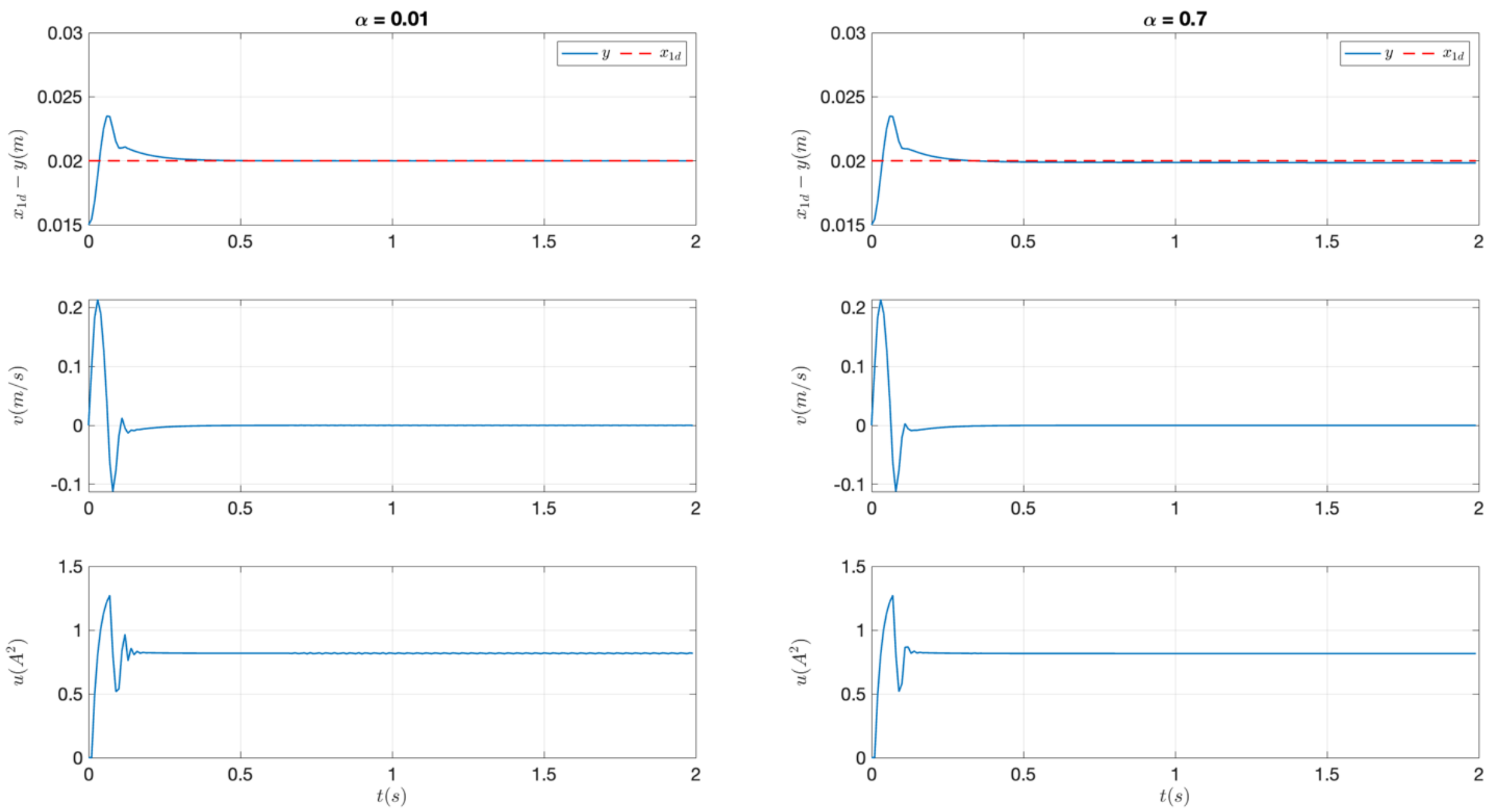

(a) (b)

**Fig. 2** Changes in the vertical position, velocity of the ball, the controller signal of the coil for ball set-point regulation with $x_{1d} = 0.02$, (a) $\alpha = 0.01$ is selected, (b) $\alpha = 0.7$ is selected

Simulation results are performed for two different purposes, set point regulation, and time-varying trajectory tracking of ball position in Maglev system. The ball references in the set point regulation and time-varying trajectory tracking are considered as $x_{1d} = 0.02$ and $x_{1d} = 0.02 + 0.01\sin(2\pi t)$, respectively. In addition to that, the simulation results are presented for the analysis of the Caputo derivative selection with $\alpha = 0.01$ and $\alpha = 0.7$. The controller design parameters in the simulation cases for $\alpha = 0.01$ and $\alpha = 0.7$ are assigned as $k_1 = 50$, $k_2 = 100$, $k_3 = 0.01$ and $k_4 = 1$.

In the numerical simulation results, the changes in the vertical position, velocity of the ball, the controller signal of the coil for the ball set-point regulation in Fig. 2 and time-varying trajectory tracking in Fig. 3 are presented. It is observed from Fig. 2a that the ball converges successfully to set point value in 0.5 s. However, in Fig. 2b, there is a steady-state error in the ball position at a negligible level. Moreover, Fig. 3 shows the performance of the time-varying trajectory tracking of the magnetic ball. It can be seen that the results of steady state performance are close to each other in both simulations, but it is noticed that the oscillations are different in the transient response. Additionally, it is observed that the fluctuations in both the speed and the produced control signal are more numerous for $\alpha = 0.7$ due to this oscillation.

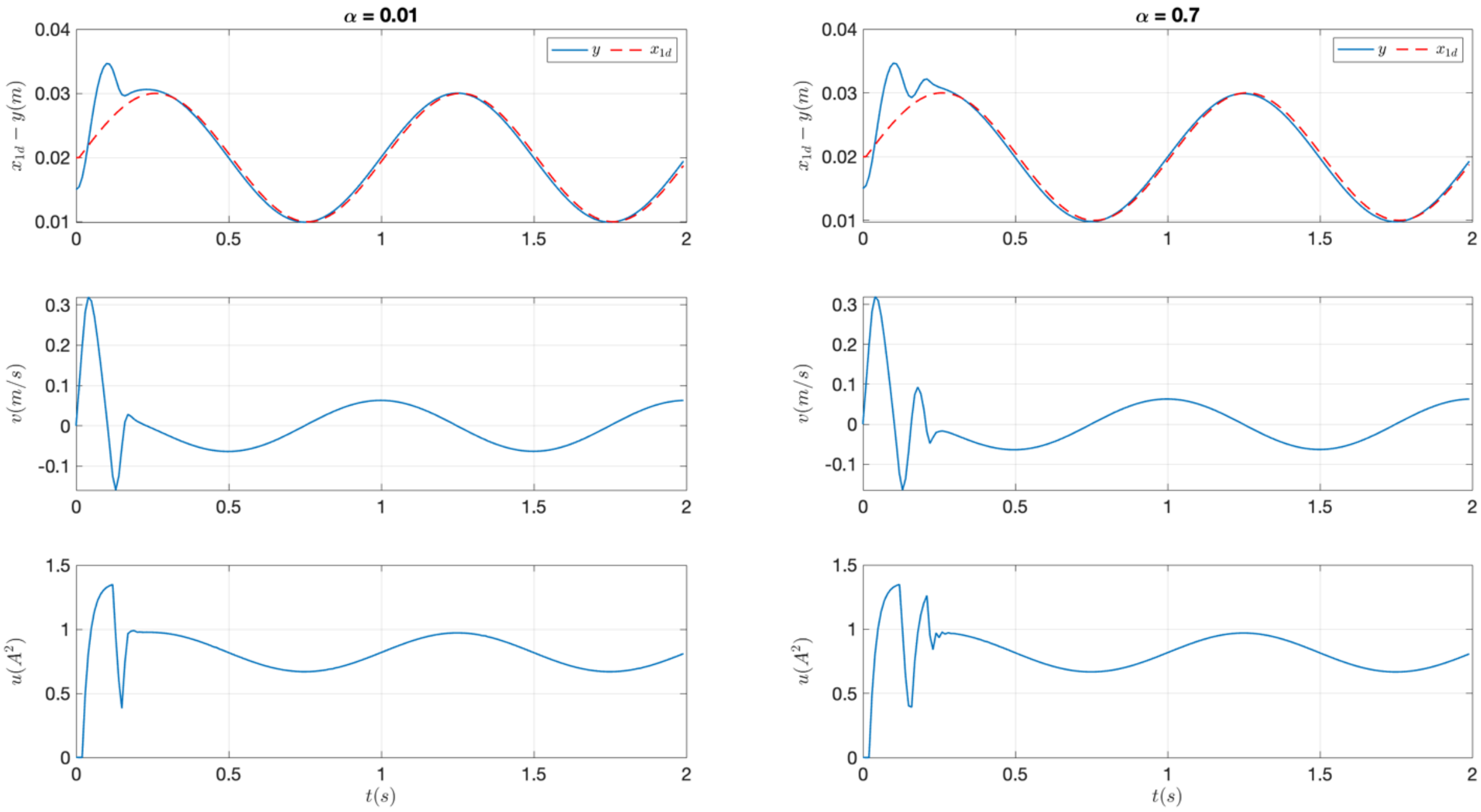

(a) (b)

**Fig. 3** Changes in the vertical position, velocity of the ball, the controller signal of the coil for ball time-varying trajectory tracking with $x_{1d} = 0.02 + 0.01\sin(2\pi t)$, (a) $\alpha = 0.01$ is selected, (b) $\alpha = 0.7$ is selected

## 5. Conclusions

In this paper, a nonlinear backstepping controller with fractional-order calculation is proposed to control the ball position of the current-controlled magnetic levitation system. The mathematical formulation of the Maglev system is presented, and the structure of the proposed fractional-order controller is devised by operating the Caputo derivative in the steps of backstepping. The asymptotic stability in the sense of Lyapunov is shown in the system dynamics under the nonlinear backstepping controller including the fractional-order calculations. The computer-based numerical simulation results validate the effectiveness and validity of the fractional-order backstepping for the Maglev system.